\documentclass{article}

\usepackage{microtype}
\usepackage{graphicx}
\usepackage{subfigure}
\usepackage{booktabs} % for professional tables
\usepackage[page]{appendix} % print appendices title
\usepackage[colorinlistoftodos]{todonotes}
\usepackage{amsmath}
\usepackage[numbers]{natbib}
\usepackage{bm}
\usepackage{float}

\usepackage{hyperref}

\usepackage[accepted]{icml2020}

\icmltitlerunning{Evading Malware Classifiers via Monte Carlo Mutant Feature Discovery}

\begin{document}

% Space before and after the equations
\setlength{\abovedisplayskip}{3.5pt}
\setlength{\belowdisplayskip}{3.5pt}

\twocolumn[
\icmltitle{Evading Malware Classifiers \\via Monte Carlo Mutant Feature Discovery}

% It is OKAY to include author information, even for blind
% submissions: the style file will automatically remove it for you
% unless you've provided the [accepted] option to the icml2020
% package.

% List of affiliations: The first argument should be a (short)
% identifier you will use later to specify author affiliations
% Academic affiliations should list Department, University, City, Region, Country
% Industry affiliations should list Company, City, Region, Country

% You can specify symbols, otherwise they are numbered in order.
% Ideally, you should not use this facility. Affiliations will be numbered
% in order of appearance and this is the preferred way.
\icmlsetsymbol{equal}{*}

\begin{icmlauthorlist}
\icmlauthor{John Boutsikas}{equal,ed}
\icmlauthor{Maksim E. Eren}{equal,ed}
\icmlauthor{Charles Varga}{ed}
\icmlauthor{Edward Raff}{booz,ed}
\icmlauthor{Cynthia Matuszek}{ed}
\icmlauthor{Charles Nicholas}{ed}

\end{icmlauthorlist}

\icmlaffiliation{ed}{Department of Computer Science and Electrical Engineering,
University of Maryland Baltimore County, 1000 Hilltop Circle, Baltimore, MD, 21250.}
\icmlaffiliation{booz}{Booz Allen Hamilton}

\icmlcorrespondingauthor{John Boutsikas}{iboutsi1@umbc.edu}
\icmlcorrespondingauthor{Maksim E. Eren}{meren1@umbc.edu}

% You may provide any keywords that you
% find helpful for describing your paper; these are used to populate
% the "keywords" metadata in the PDF but will not be shown in the document
\icmlkeywords{Machine Learning, ICML}

\vskip 0.3in]

% this must go after the closing bracket ] following \twocolumn[ ...

% This command actually creates the footnote in the first column
% listing the affiliations and the copyright notice.
% The command takes one argument, which is text to display at the start of the footnote.
% The \icmlEqualContribution command is standard text for equal contribution.
% Remove it (just {}) if you do not need this facility.

%\printAffiliationsAndNotice{}% leave blank if no need to mention equal contribution
\printAffiliationsAndNotice{\icmlEqualContribution} % otherwise use the standard text.

\begin{abstract}
The use of Machine Learning has become a significant part of malware detection efforts due to the influx of new malware, an ever changing threat landscape, and the ability of Machine Learning methods to discover meaningful distinctions between malicious and benign software.
Antivirus vendors have also begun to widely utilize malware classifiers based on dynamic and static malware analysis features.
Therefore, a malware author might make evasive binary modifications against Machine Learning models as part of the malware development life cycle to execute an attack successfully.
This makes the studying of possible classifier evasion strategies an essential part of cyber defense against malice.
To this extent, we stage a grey box setup to analyze a scenario where the malware author does not know the target classifier algorithm, and does not have access to decisions made by the classifier, but knows the features used in training. 
In this experiment, a malicious actor trains a surrogate model using the EMBER-2018 dataset to discover binary mutations that cause an instance to be misclassified via a Monte Carlo tree search. Then, mutated malware is sent to the victim model that takes the place of an antivirus API to test whether it can evade detection.

\end{abstract}

\section{Introduction}
\label{introduction}

The number of malware in the wild has increased significantly in recent years. 1.1 billion total malware was recorded in 2020 alone. That is an over 139\% increase from 2015 \cite{avtest_2020}. 
The influx in the quantity of malware makes Machine Learning (ML) approaches such as statistical modeling, behavior analysis, and deep learning an ideal choice for malware analysis by cyber defenders. 
When used as a helper in the cyber domain, ML has shown to be one of the most effective ways to reduce risk, drive precise detection, reduce cost, and reduce response and recovery time \cite{accenture_innovate}. 
Because of its advantages, ML and Artificial Intelligence (AI) are utilized by 38\% of organizations \cite{accenture_cost}, and 83\% of these spend more than 20\% of their cybersecurity budget on such technologies \cite{accenture_innovate}. 
At the same time, antivirus (AV) vendors have begun to widely utilize ML based malware detection techniques \cite{WinIntel, VirusTotalSangfor, VirusTotalBitdefender, Fleshman2018}. 

Recent work has shown that top AV that utilize some form of ML technology can be bypassed with simple modifications on the malware such as by adding a new section, appending a single byte, removing the debug and certificate values, or renaming a section \cite{Song2020, anderson2018learning}.
In our work, we borrow the feature changes that are shown to be effective in prior research, and explore a new mutant malware discovery methodology that is based on Monte Carlo Tree Search (MCTS).

We approach classifier evasion as a game playing exercise between the adversary and the ML model where a winning hand is a successful mutation that makes the malware undetectable. Similar to chess, there are numerous state permutations -- mutations in our case -- that can yield a winning play at any stage of our "game".
MCTS can find the winning hand by simulating a subset of all possible mutations, and discover an optimal path using an empirical scoring method. This allows empirical evaluation of mutations while minimizing error, and examines paths without actually computing all the possible permutations of changes.
At the end, MCTS prioritizes minimizing the number of trackable changes that lead to evasion, hence avoiding excessive changes to the binary.

In this experiment, we stage a grey box adversarial scenario where the target classifier algorithm, decisions made by the classifier, and the data used in training are unknown, but the features used in training are known. The attacker trains a local Decision Tree (DT) with the test portion of the EMBER-2018 dataset \cite{Anderson2018} and discovers evasive feature modifications via MCTS while using the surrogate model (DT) for confirmation. 
This allows attackers to circumvent querying AV APIs to verify their modifications. 
Such a scenario is feasible for an adversary when an API for verification is unavailable because it is part of an internal defense system at an organization, or if the malicious actor wants to avoid attention.
The performance of the mutations is then evaluated against the victim Multilayer Perceptron (MLP), which is trained on the training portion of EMBER-2018, that takes the place of the target antivirus engine.

Our results show that MCTS finds successful mutations on the surrogate model for 52.18\% of the malware, out of which 8.78\% evades detection by the victim model, performing slightly better than our random mutation baseline with a 5.26\% evasion rate. 
%In future work, we plan to explore MCTS performance on varying families of surrogate and victim ML models. 
Our contributions in this paper include:
\vspace{-3pt}
\begin{itemize}
\setlength\itemsep{-0.3em} % space between each item
\item Demonstrating an evasive malware feature discovery technique with Monte Carlo tree search. 
\item Analyzing the detection avoidance success rate when the adversary does not have access to the ground truth via an AV API.
\end{itemize}

\section{Relevant Work}
\label{sec:background}

\begin{table*}[t!]
\caption{Statistics for 299,867 benign and 299,851 malicious binaries in EMBER-2018 training set} \label{table:training_set_stats}
\centering
\resizebox{\columnwidth * 2}{!}{
\begin{tabular}{@{}lcccccccc@{}}
\toprule 
            \multicolumn{4}{c}{\hspace{5.6cm}Malware}                  		& \multicolumn{4}{c}{\hspace{4.5cm}Benign Software}       \\ \cmidrule(l){2-5} \cmidrule(l){6-9} 
                    & \textbf{Mean}     & \textbf{Std}      & \textbf{Min}   & \textbf{Max}            	& \textbf{Mean}         & \textbf{Std}      & \textbf{Min}   & \textbf{Max}     \\ \midrule
Strings Entropy     & 5.967             & 0.615             & 0     & 6.584          		& 5.595      & 0.659      & 0     & 6.585       \\
Number of Strings   & $6.12\times 10^3$ & $1.67\times 10^4$ & 0     & $1.63\times 10^6$     & $8.26\times 10^3$      & $3.37\times 10^4$      & 0     & $2.48\times 10^6$       \\  
File Size           & $1.24\times 10^6$ & $2.39\times 10^6$ & 512   & $2.71\times 10^8$ 	& $1.71\times 10^6$      & $6.93\times 10^6$      & $2.34\times 10^2$     & $5.36\times 10^8$       \\
Number of Exports   & 9.019             & 166.313           & 0     & $5.26\times 10^4$     & 51.877      & 625.764      & 0     & 52628       \\
Number of Imports   & 98.993            & 140.209           & 0     & 3074       			& 113.456      & 286.144      & 0     & 21344       \\ 
Timestamp (POSIX)           & 1358407000        & 416879700         & 0     & 4294967000      		& 1332756000      & 574590400      & 0     & 4294967000       \\
Size of Code        & $2.75\times 10^6$ & $7.10\times 10^7$ & 0     & $4.29\times 10^9$     & $5.92\times 10^5$      & $4.17\times 10^6$      & 0     & $1.67\times 10^9$       \\
Number of Sections  & 5.088             & 3.229             & 1     & 97      				& 4.854      & 2.639      & 0     & 198       \\\bottomrule
\end{tabular}
}
\end{table*}

Adversarial attacks against ML models and defenses for these attacks are widely studied research areas \cite{jia2020entangled, papernot2017practical, tramer2017ensemble}. Consequently, 
an open source toolbox for adversarial training has become popular in recent years \footnote{\url{https://github.com/Trusted-AI/adversarial-robustness-toolbox}}. 
These attacks include modifications on the training data, poisoning the feature vector, or extracting the ML model parameters.
When it comes to malware data science, however, there are special considerations to be taken.

Song et al. argues that some attacks are not always realistic or have an effect on the end user in the malware domain \cite{Song2020}. 
Attacks performed on feature vectors are not realistic as the poisoned feature vector may not possess the corresponding binary representation. Conversely, the attacks on byte-based models like MalConv \cite{Raff2020b,MalConv} insert adversarial content into an executable that maintains validity while attempting to hide the evasive payload \cite{Suciu2019,Kolosnjaji2018,Kreuk2018}. An additional constraint (to the defender's advantage) is that attacks are one-way, as benign authors do not attempt to alter their files to become falsely detected as malicious -- only malicious authors alter their files to avoid detection \cite{Fleshman2018a,Incer:2018:ARM:3180445.3180449}. 
Therefore, in our analysis we avoid performing attacks towards feature vectors and focus on modifying the malware itself.

Similarly, changing malware requires additional ruminations as the direct modifications of the compiled malware may make the binary non-functional \cite{Fleshman2018,Song2020}.
Malware that is generated in such a manner can evade the detector, but at the same time it may no longer demonstrate the malicious behavior. 
A \textit{keylogger} could, for example, obtain permanence and display some of its malware characteristics, but may no longer actively log keystrokes, as that portion of the code has been altered during the modifications to avoid identification. 
We recognize this problem by developing heuristics that limit the possible changes in the Monte Carlo Tree Search. 
With that, we introduce a strong assumption that after the modification(s), the malware is still functional. 
In a real world example, an adversary can alter this rule set if the changes break the malware. 
Nevertheless, malware functionality can be verified on a virtual sandbox environment, and we leave this significant step to future work.

Additionally, Song et al. points out that some of the attacks demonstrated by other research assume inside knowledge with a white box setup, which is not realistic for an adversary to possess in real world scenarios \cite{Song2020}.
For example, an attacker may not have access to the malware classifier's architecture when it is part of the internal company infrastructure, unless the attacker obtains such knowledge via other means in the reconnaissance stage of the attack. 
Furthermore, in such a scenario, the attacker does not have access to the decisions the malware classifier makes. 
With this in mind, we stage a grey box attack where it is assumed that the attacker lacks access to the malware classifier but knows the features used in training; hence, the attack must be conducted without the ground truth in classification. 
Consequently, in our setup an adversary trains a surrogate model using a different set of data to verify the success of mutations.

With all the important considerations above, Song et al. develop a malware modification and mutant malware functionality verification system, and show that the static classifiers of real AV programs can be evaded 24.3\% to 41.9\% of the time with changes as simple as a single byte modification \cite{Song2020}. 
Finally, they discuss the fact that even though a sample can evade the classifier, it is not known what factors contribute to this evasion. 
In order to alleviate this concern, they provide a framework that can generate malware mutations while preserving the original malicious behavior of the sample, and also providing a way to trace back those modifications in order
to explain which features were responsible for the evasion.
Similarly, our mutant feature searching method can re-track the changes that caused the original mutation with a minimal number of steps.

\section{Dataset}
\label{sec:dataset}

%\begin{figure*}
%\begin{center}
%    \includegraphics[width=\columnwidth]{figures/roc_mlp.png}
%    \includegraphics[width=\columnwidth]{figures/pr_mlp.png}
%\end{center}
%\small\normalsize
%\begin{quote}
%\caption[ROC-AUC]{
%    Left: ROC curve and the AUC for the victim Multilayer Perceptron (MLP) and
%    the surrogate Decision Tree (DT) before and after the mutations. 
%    Right: PR curve and F1 scores for the victim Multilayer Perceptron (MLP) and surrogate Decision Tree (DT) before and after the mutations.
%}
%\label{fig:roc-auc}
%\end{quote}
%\vskip -0.3in
%\end{figure*}
%\small\normalsize
%\raggedbottom

We base our analysis on the EMBER-2018 dataset \cite{Anderson2018}. 
EMBER contains features extracted from 1.1M Windows PE files in a human friendly, JSON format, along with their label indicating whether the sample is malicious or benign.
The main reason for our choice is the same that drove the creation of EMBER. 
A large collection of binaries is bound by many difficulties including logistical ones (i.e. storage, safe distribution) and legal ones (i.e. copyrighted binaries). 
EMBER resolves that by not sharing the binaries themselves -- or even their name for that matter -- but by sharing a parsed representation of various metadata that can be extracted from the PE header. We use this metadata to train our surrogate and victim models.

The default training and test set split in the EMBER dataset has a significant factor in our grey-box setup. 
Our victim malware classifier is trained on the EMBER-2018 training portion which consists of 299,867 benign and 299,851 malware instances for a total of 599,718 samples. On the other hand, our surrogate (adversary's local model) will only utilize the EMBER-2018 test set that contains 199,800 samples, of which exactly half are malicious.
This split supports our grey-box setup as in the real world the attacker would not have access to the malware samples used in training of the target malware classifier, and would most likely have access to fewer samples than the  entity supporting the target classifier.
In Table \ref{table:training_set_stats} we compare the statistics for malware and benign-ware in the EMBER training set.

At first glance, we see that malware has a higher mean string entropy compared to benign software on average, as a result of malware often being obfuscated to avoid detection. Obfuscation randomizes the information in malware and increases the entropy value. Meanwhile, the average number of library exports and imports is greater for benign software. This supports the idea that malware is developed for more specific tasks requiring a smaller number of imports and exports. We then proceed to the pre-processing of our dataset, utilizing the tools available in the Scikit-learn library \cite{pedregosa2011scikit}.

\subsection{Pre-processing}
\label{sec:preprocessing}
During training, both the surrogate model and the victim classifier utilize numerical and categorical static malware analysis based features. 
The numerical features used in training are \textit{String Entropy}, \textit{Number of Strings}, \textit{File Size}, \textit{Timestamp}, \textit{Size of Code}, \textit{Number of Sections}, \textit{Number of Exports}, and \textit{Number of Imports}. These features are scaled to unit variance during pre-processing, and missing values are replaced with the median. 

Categorical attributes include the \textit{Has Debug} flag which indicates if the binary has the debug value set, the \textit{Has Signature} which is used to indicate the existence of the binary certificate, and the \textit{Entry} string that names the section name the binary uses to start the code execution. Categorical features are vectorized via one-hot encoding and missing values are replaced with the most frequent element for each feature. Unknown categorical attributes in the validation set are ignored. We also used \textit{imported libraries} and \textit{imported functions} during training. Unlike the previous categorical features, \textit{imported libraries} and \textit{imported functions} are vectorized by hashing their values while limiting the number of dimensions to $2^{10}$ for libraries and $2^{13}$ for the functions. Finally, all the features that are mentioned above are combined via multimodal feature fusion to create a single features vector \textbf{$X$} representing each $n$ malware sample.

\section{Adversarial Model}
\label{sec:adversarial_model}

Adversarial scenarios differ in terms of the knowledge that an attacker possesses about the target. 
In a white-box adversarial setting, it is assumed that the attacker has significant knowledge about the target model, including the algorithm and the features-set used in training.
On the other hand, black-box attacks are more realistic as the only known information is the output label of an otherwise unknown algorithm.
For the malware classifiers; however, the output label may also be unavailable when the model is part of an internal defense system at an organization. 
Therefore, an attacker needs to verify the validity of modifications locally before the attack is deployed.

Our adversarial setup follows a grey-box scenario, a structure that is in between the white-box and the black-box setting, where the malware features and their corresponding pre-processing steps are the same for the surrogate and the target model. All the other information about the victim model is assumed to be unknown, including the prediction labels. To this extent, we train two different models where one takes the place of the victim malware classifier an AV would have used, and a separate surrogate classifier that the malicious actor trains to identify the evasive feature mutations. 
In this section we describe both of these models, and provide their performance with receiver operation characteristics (ROC) and area under the curve (AUC), and with a precision-recall (PR) curve and the F1 scores.

\subsection{Grey-box Surrogate Learner}

Because the malicious actor does not have access to the decisions made by the target classifier, a local Decision Tree is trained, using Scikit-learn \cite{pedregosa2011scikit}, on a random subset of 60\% of the EMBER test set. Predictions made by the Decision Tree are used to identify the features which makes the mutation valid. A valid mutation is when a malware sample is predicted to be benign by the surrogate after a mutation or combination of mutations.

%A Decision Tree is a supervised learning model where the feature space is recursively partitioned on each iteration by choosing the feature which produces the purest sub-trees during training \cite{Aurelien2019}. This is done with the Classification and Regression Tree (CART) algorithm with the following cost function:

%\begin{align*}
%J(d, t_d) = \dfrac{n_{left}}{n} \cdot G_{left} + \dfrac{n_{right}}{n}\cdot G_{right}
%\end{align*}
%where $G_{left}$ and $G_{right}$ measures the impurity of the left and right subset, $n_{left}$ and $n_{right}$ are the number of instances belonging to the left and right sub-tree, and $t_d$ is the purity threshold for feature $d \in \{1, 2, 3, \ldots, D\}$ \cite{Aurelien2019}. The training is completed when the algorithm can no longer split the features or when the tree reaches to the hyper-parameter depth of 12, and the leaf is used as the prediction label.

We set the Decision Tree (surrogate) maximum depth hyper-parameter to 12. Our model obtains an ROC-AUC of 0.956 and an F1 score of 0.916 when classifying the remaining 40\% of the EMBER-2018 test set.

\subsection{Victim Malware Classifier API}
\label{subsec:api}

% PR and ROC curves of MLP 
%\begin{figure}[htb]
%\centering
%\begin{minipage}[b]{.45\textwidth}
%\centerline{\includegraphics[width=\columnwidth]{figures/roc_mlp.png}}
%\caption{ROC curve and the AUC for the Multi-later Perceptron (MLP) and Decision Tree (DT) models.}\label{roc-auc}
%\end{minipage}\qquad
%\begin{minipage}[b]{.45\textwidth}
%\centerline{\includegraphics[width=\columnwidth]{figures/pr_mlp.png}}
%\caption{PR curve and F1 scores for the Multi-later Perceptron (MLP) and Decision Tree (DT) models.}\label{pr-auc}
%\end{minipage}
%\end{figure}

We train a Multi-layer Perceptron (MLP) that takes the place of the AV. We train our model on the EMBER training set with a 20\% split for the validation set which is used in hyper-parameter tuning. 
Our model consists of seven hidden fully connected layers with 512, 256, 128, 256, 128, 256, and 128 hidden nodes, respectively. In the hidden layers of the MLP we use the Rectified Linear Unit (ReLU) activation function \cite{nwankpa2018activation, Nair2010RectifiedLU}. The output layer with a single node transfers the input $x_i$ with the Sigmoid activation function $\dfrac{1}{1 + e^{-x_i}}$. For the model optimizer, we use Adaptive Moment Estimation (Adam) \cite{kingma2019method}. Adam's exponential decay hyper-parameters $\beta_1$ and $\beta_2$ are left at default $\beta_1 = 0.9$ and $\beta_2 = 0.999$ as suggested in \cite{kingma2019method}. 

With this setup, our model obtains the ROC-AUC score of 0.964 and F1 score of 0.934 when classifying the 200,000 benign and malicious instances in the held out EMBER-2018 test.

\section{Mutant Malware Generation}
\label{sec:malware_modification}

As discussed earlier, our goal is to generate mutations of the malware in the
EMBER-2018 dataset, such that the mutated malware will not be detected by the API
classifier. However, it is very important that those samples remain malicious
after the modification. To that end, we provide a way to limit the mutations
that can be applied to any given sample -- mutated or otherwise. We will next
explain these limitations and their purpose. There is a difference in the complexity of the
mutations, as some imply that a number of different, simpler mutations also
happen, and can be thought of as aggregate mutations.

\subsection{Target Features}
\label{sec:malware_modification_rules_features}
Here we list the set of possible mutations available to our Monte Carlo
implementation, and their corresponding heuristic that limits the changes:

\begin{enumerate}
\item \textbf{Add String}: Any sample can have a total of 15 strings added
to it, the reason being that there is a limit to the free, unused space
in an already compiled binary, and as such there is a limit to how many
strings can be added by using that space.
\item \textbf{Add String with Size}: This mutation shares the same 15
strings limit as the previous mutation; however, only 5 of these strings can
also modify the size of the binary. The purpose of this simulation is to
account for strings that can be added by small extensions to existing
sections of the binary. Our implementation changes the size of the binary by
30 bytes.
\item \textbf{Change String Entropy}: This mutation attempts to modify the
string entropy of the binary, towards a "benign" value. For our benign value
we use the average entropy of all benign samples. There can be a total of 7
entropy modifications per sample. It represents adding specific strings with
the goal of modifying the entropy towards a target value, versus just adding
any string like the previous 2 mutations did.
\item \textbf{Change String Entropy with Size}: This mutation is similar to
the previous one, and shares the same 7 applications cap. However, only 3 of
those 7 mutations can also affect the size of the binary. The major
difference with this mutation, is that since the attacker can specify the
size, it is easier to construct strings that provide a more significant
change to the overall entropy. Both entropy mutations imply the
addition of a string.
\item \textbf{Remove String}: By removing a string, we are trying to
simulate the act of the attacker finding a series of random bytes that are
interpreted as strings, but they are in fact not strings, and changing those bytes so
they are no longer detected as strings by the feature extraction. We limit
these removals to 4 per sample, as those type of strings should be less
frequent.
\item \textbf{Add Section}: With this mutation we add a section with benign
contents to the sample. There is no limit to how many times this can be
applied, as after some testing we found that our algorithm would avoid
adding a large quantity of sections, and instead use a shorter route to
evasion. Each section adds 512 bytes to the sample's size.
\item \textbf{Add Bytes}: In this mutation we simulate appending benign
bytes to any of the existing malware sections. This mutation adds 128 bytes
to the size of that section and the overall size of the file. The mutation
can be applied as many times as desired.
\item \textbf{Add Code Bytes}: Appends 64 benign bytes to the code section
specifically. This mutation be applied as many times as desired, but again our algorithm
seems to avoid extensive application of these.
\item \textbf{Import Function}: This mutation will add a function to the
import table of the malware, as well as the matching DLL if it is missing.
In order to decide what those functions would be, we made a list of the most
common functions in the benign samples that do not appear in malware
samples. This gave us 14 candidate functions, so each application of this
mutation will select one of those 14 at random, provided it is not already
present in the sample.
\item \textbf{Change Timestamp}: This mutation will modify the timestamp of
the sample towards a target timestamp, by the given step size. For our
target timestamp we used the average timestamp of all benign samples, with a
step of 1000 milliseconds. The mutation can be applied as many times as desired, as it can
adjust the direction of the step to move towards the target timestamp.
\item \textbf{Remove Debug}: This mutation will set the debug flag to false.
The mutation can only be applied if the sample's debug flag is set to true.
\item \textbf{Change Signature}: This mutation sets the certificate flag to
true. The mutation can only be applied if the sample does not have a
certificate already.
\end{enumerate} 

We use this list of feature modifications as they seemed
to provide reasonable results and prior work showed that successful evasion is possible
with some of them \cite{Song2020}. Our framework allows for easy modification of
the mutation values -- by the researcher -- to match any given dataset, including
restricting the heuristics if the modifications break the malware. It also
allows for easy extension of these mutations to accommodate for the samples at
hand.

\subsection{Monte Carlo Mutant Feature Search}
We decided to approach the mutation finding activity as a game playing problem
where a winning hand is a successful mutation that makes the sample
undetectable. This fits well with our problem because at any stage of our
"game", there are many different possible combinations of mutations that can be
the winning hand. There are multiple algorithms that can be used in this context,
including, but not limited to, the \textit{mini-max alpha beta pruning} and the
\textit{expectimax tree search} algorithms. The main drawback of these
algorithms is that they rely on the existence of two opponents playing the game.
Thus, our analysis direction changed towards a game playing algorithm
that could be adapted to a "single player" context as our only player would be
the program that is trying to beat the surrogate model (i.e. the surrogate model cannot
make any "plays"). As such, we decided to proceed with the \textit{Monte Carlo
Tree Search} (MCTS) algorithm. This algorithm still shares all the tree-related
pitfalls -- like state explosion and extreme redundancy -- with the previously
mentioned algorithms. We address those issues in our
implementation.

MCTS has gained popularity based on its application for the game \textit{Go} as
described by \cite{coulom2009monte}, as well as its latest implementation in
Google's AlphaZero \cite{Silver1140alphazero}. The algorithm consists of 4
conceptual stages. All stages happen once per iteration.

\begin{enumerate}
\item \textbf{Tree Traversal}: In this phase, the algorithm traverses the
existing nodes of the tree. A node is a combination of the mutations that
have happened to the sample up to that point. The goal is to reach a node
that is currently a leaf, and expand it if it is not a terminal node. We
consider a node terminal when a benign classification is reached. We name
this behavior the \textit{Tree Policy}.
\item \textbf{Expansion}: During expansion the algorithm will add a number
of children nodes to the node found during tree traversal. Each child will
represent a mutation allowed by our rules. The expansion of a given node
happens only once; the second time a non-terminal, leaf node is visited. The
first time we visit a leaf node, we simulate it on its own; therefore, we
cannot also expand it at the same time. We name this the
\textit{Expansion Policy}.
\item \textbf{Simulation}: The application of Monte Carlo methods happens in
this phase. Starting from the node that is currently evaluated, the
algorithm generates subsequent states at random until it reaches a terminal
node or we reach our simulation depth. At the end of this phase, a score is
calculated for the node that is under evaluation by accounting for the simulated
number of mutations, then the simulation states are discarded. Finally, the score is
then propagated back up the tree. We name this behavior the
\textit{Simulation Policy}.

\item \textbf{Back propagation}: In this phase, we update the score
of the node we just evaluated as well as its visit count, and then propagate
this change up the tree until we reach the root node. This way the tree
policy can select a more suitable path to explore in the next iteration.
\end{enumerate}

MCTS on its own does not suffice. Even successful implementations --
AlphaGo Zero for instance -- do not rely on the
algorithm alone, and make modifications. Similarly, we implement the
following modifications to the various phases of MCTS, to better suit our
problem.

We borrow inspiration from the Upper Confidence Bounds algorithm as applied to
trees (UCT) for the \textit{Tree Traversal} phase \cite{Kocsis06banditbased}.
UCT allows selecting the next best child to explore using an empirical score,
through the use of bandit methods, instead of relying on obtaining the actual
score. Computing the real score of the child is expensive, and quite often
impossible for our case as finding a successful mutation from any given node is
not always achievable. However, the use of the empirical score introduces a
potential divergence from the true value. Thankfully, the algorithm is shown in
the original paper to be consistent within the estimation error caused by
sampling. We apply a modification on that algorithm, and use the UCB1
function to evaluate the next child to traverse:

\[
\textit{ucb1} = \frac{\mathit{score_{child}}}{\mathit{visits_{child}}} + c\cdot \sqrt{\frac{\ln(\mathit{visits_{parent}})}{\mathit{visits_{child}}}}
\]

where the first element in the sum is the empirical score of the child node we
want to evaluate, $c$ is the exploration coefficient, and the squared root
factor gives a higher score to the children that we have visited the least. In
this context, a 0 visit means always exploring the child node. Finally, a high
exploration coefficient makes our traversal more akin to breadth-first search,
while a lower value makes the traversal similar to a depth-first search. We put
these modifications in place with the hope that our search will not limit itself
into exploring a few specific mutations, but will instead explore a larger
portion of the mutation space.

By using the above evaluation score the \textit{Tree Policy} selects the child
with the highest evaluation to traverse next with:

\[
\mathit{next\_child} = \operatorname*{arg\,max}_{ucb1} \; \mathit{children}
\]

This allows our search to change which part of the tree it explores in each
iteration, and give high prioritization to the nodes that have not been explored
before using the UCB1 based evaluation.

\begin{algorithm}[!ht]
\caption{Tree Policy}
\label{alg:tree_policy}
 \begin{algorithmic}
\STATE {\bfseries Input:} node $root$
\STATE Initialize $node = root$
\WHILE {$node.is\_expanded$}
\STATE $node = \operatorname*{arg\,max}_{ucb1} node.children$ 
\ENDWHILE
 \end{algorithmic}
\end{algorithm}

Similarly, we apply some modifications in the expansion phase. We begin with our
mutations rules -- or invariants -- mentioned in Section \ref{sec:malware_modification_rules_features}.
In the original MCTS, a child is added for each potential action from the current
node when a node is expanded. Specifically, we limit the available children
nodes to the ones with applicable mutations according to our invariants. This
reduces the potential state explosion, and limits generating mutations that are
likely to break the malware. Finally, we need to address the redundancy present
in a tree structure. If $0, 1, 4$ are 3 different mutations, the classification
result for $[0, 4, 1]$ and $[4, 1, 0]$ should be the same since we examine the
mutation set as a whole instead of each mutation individually. By sorting the
proposed path and hashing it, we can cross-reference the hash of the sorted new
mutation with all the other, sorted paths which we have already examined. If we
have a match for this proposed path for the malware sample $n$, the mutation is
not added during the expansion phase as a child node. This further reduces the
potential state expansion as well as our runtime, as we no longer spend time on
the previously seen combinations.

\begin{algorithm}[!ht]
\caption{Expansion Policy}
\label{alg:expansion_policy}
 \begin{algorithmic}
\STATE {\bfseries Input:} Current node $node$
\STATE {\bfseries Input:} Existing mutations $mutations$
\STATE {\bfseries Input:} Seen paths $seen\_paths$

\FOR {$mutation \in mutations$}
\IF {$mutation.is\_allowed$}
\STATE Proposed path
\STATE {$p$ = $node.mutation\_path \cup mutation$}
\STATE Hashed path
\STATE {$h = hash(sort(p))$}
\IF {$h \notin seen\_paths$}
\STATE Make a new child based on $mutation$
\STATE {$node.children = node.children \cup new\_child$}
\ENDIF
\ENDIF
\ENDFOR
 \end{algorithmic}
\end{algorithm}

Our simulation phase is not that different from the original algorithm. However,
we do still enforce our invariants when expanding the randomly selected nodes.
If we skip this step, the algorithm could generate a successful mutation that is
not allowed by our invariants. Therefore, this part of the tree would get a high
score, but the search would never be able to construct the path as it is not
considered valid during the expansion phase. During the simulation phase, child
selection happens through random sampling instead of UCB1 evaluation.

\begin{algorithm}[!ht]
\caption{Monte-Carlo Tree Search}
\label{alg:search_pseudo}
 \begin{algorithmic}
\STATE {\bfseries Input:} sample $s$, iterations $it$
\STATE Initialize $iterations = 0$.
\STATE Initialize $root$ from sample
\STATE $node = root$
\REPEAT
\WHILE {$node.is\_expanded$}
\STATE $node$ = TreePolicy(node)
\ENDWHILE
\IF {$node.visit\_count \neq 0$}
\STATE ExpansionPolicy($node$)
\STATE $node$ = TreePolicy(node)
\ENDIF

\STATE $path$ = SimulationPolicy($node$)
\STATE $score$ = EvaluatePath($path$)
\STATE BackPropagate($score$)
\UNTIL{$iterations$ is $it$}
 \end{algorithmic}
\end{algorithm}

Finally, before we move onto back propagation, we need a value to assign as the
score of the newly evaluated node. We elected to use the length of the simulated
path to a benign mutation. So for example, if our simulation policy produced the
path $[7, 8, 4, 4, 9, 10]$ from some node $n$, then the score of $n$ will be
$length(path) = 6$. However, since we care about the shortest mutation paths we
use the negative of that score. This allows the shortest paths to produce
the higher scores. If no benign classification was found, we set the score to $
- \infty$:
\begin{equation}
\label{eq:score_eval}
s = \left\{
\begin{array}{@{}ll@{}}
- length(min(path)), & \text{if}\ \text{benign} \\
- \infty, & \text{otherwise}
\end{array}\right.
\end{equation} 
We should note that this score is produced only once, as each node is simulated
only once. It can be, however, updated since in the subsequent iterations the search can
explore the children of $n$. The back propagation phase of these children will
modify the score of $n$.

\begin{algorithm}[!ht]
\caption{Back propagation}
\label{alg:back_propagation}
 \begin{algorithmic}
\STATE {\bfseries Input:} Node's ancestors $ancestors$
\STATE {\bfseries Input:} Score $s$

\FOR { $node \in ancestors$}
\IF {$node.score \neq -\infty \textbf{\ AND\ } s \neq -\infty$}
\STATE {$node.score \mathrel{+}= s$}
\ELSE
\STATE {$node.score = s$}
\ENDIF
\STATE {$node.visits \mathrel{+}= 1$}
\ENDFOR
 \end{algorithmic}
\end{algorithm}

We provide pseudocode for the search in Algorithm \ref{alg:search_pseudo},
where the various policies are implementations of what we have discussed so far,
and \textit{EvaluatePath} is shown in Equation \ref{eq:score_eval}.

 \begin{algorithm}[!ht]
\caption{Path recovery}
\label{alg:path_recovery}
 \begin{algorithmic}
\STATE {\bfseries Input:} node $root$
\STATE {\bfseries Output:} Mutation Path $path$
\STATE Initialize $node = root$.
\STATE Initialize $path$ = empty list
\WHILE {$node.is\_expanded\ \textbf{AND}\ \textbf{NOT}\ node.is\_terminal$}
\STATE $node = \operatorname*{arg\,max}_{score} node.children$ 
\STATE $path = path \cup node$
\ENDWHILE
 \end{algorithmic}
 \end{algorithm}
 
\begin{table*}[t!]
\caption{Statistics from mutation discovery via Monte Carlo and Random Search on the surrogate Decision Tree
(DT). The \textbf{Alone} column presents the number
of malware where one type of feature change was enough for misclassification.
\textbf{In Group} shows the number of samples where the mutation appeared in combination
with other mutations. The \textbf{Repeats} column provides the instance count where the mutation type appears more than once. \textbf{Affected
Instances} is the number of mutated malware, and \textbf{Total Occurrence} is the total number of times our tree
search yielded the mutation across all of the successfully modified malware samples. \label{table:mutation_stats}}
\centering
\resizebox{\columnwidth * 2}{!}{
\begin{tabular}{@{}lcccccccccc@{}}
\toprule
\multicolumn{5}{c}{\hspace{8cm}Mutations over DT with MCTS}& \multicolumn{5}{c}{\hspace{6cm}Mutations over DT with Random Search}\\ \cmidrule(l){2-6} \cmidrule(l){7-11} 
                                    & \textbf{Alone}& \textbf{In Group}& \textbf{Repeats}& \textbf{Affected Instances}  & \textbf{Total Occurrence} & \textbf{Alone}    & \textbf{In Group}     & \textbf{Repeats}  & \textbf{Affected Instances}   & \textbf{Total Occurrence}\\ \midrule
Add String                          & 0             & 79               & 23              & 79                           & 115                       & 0                 & 496                   & 27                & 496                           & 523 \\
Add String with Size                & 6             & 137              & 0               & 143                          & 143                       & 1                 & 4355                  & 0                 & 4356                          & 4356 \\
Change String Entropy               & 0             & 2223             & 90              & 2223                         & 2313                      & 0                 & 615                   & 30                & 615                           & 645\\
Change String Entropy with Size     & 9724          & 3223             & 5               & 12947                        & 12952                     & 1144              & 6754                  & 1                 & 7898                          & 7899\\
Remove String                       & 0             & 35               & 0               & 35                           & 35                        & 0                 & 4389                  & 510               & 4389                          & 4937 \\
Add Section with size               & 698           & 249              & 151             & 947                          & 1153                      & 186               & 4609                  & 506               & 4795                          & 5339 \\
Add Bytes                           & 0             & 33               & 0               & 33                           & 33                        & 21                & 4424                  & 508               & 4445                          & 4993 \\
Add Code Bytes                      & 0             & 31               & 2               & 31                           & 37                        & 20                & 4416                  & 481               & 4436                          & 4951\\
Import Function                     & 502           & 634              & 559             & 1136                         & 1949                      & 66                & 4451                  & 575               & 4517                          & 5128 \\
Change Timestamp                    & 0             & 32               & 1               & 32                           & 33                        & 0                 & 4465                  & 536               & 4465                          & 5036 \\
Remove Debug                        & 10            & 39               & 0               & 49                           & 49                        & 33                & 686                   & 0                 & 686                           & 686 \\
Change Signature                    & 37214         & 1238             & 0               & 38452                        & 38452                     & 4150              & 17470                 & 0                 & 17470                         & 17470 \\\bottomrule
\end{tabular}
}
\end{table*}
\raggedbottom

 The last missing piece of our algorithm is the recovery of the path once MCTS
 has finished all of its iterations. MCTS is often used to select the next
 action or state transition. Our goal, however, is to create a serializable
 version of the shortest mutation path so that we can apply it later against a
 different classifier. Therefore, we need to recover the whole mutation path.
 The algorithm is very intuitive, as our scoring ensures that the best evaluated
 nodes will have the highest -- although still negative -- scores and visit
 counts. With that, recovering the path is as simple as starting from the root
 of the tree then always selecting the child with the highest score until we
 reach a terminal node. If after this process we cannot reach a terminal node,
 it means that the search was unable to find a mutation for this sample.

\section{Results}
\label{sec:results}

Here we present the mutant feature discovery results and their performance when
evading the target classifier introduced in Section \ref{subsec:api}. Additionally, we compare the Monte Carlo search and its corresponding classifier evasion
performance to a Random Search baseline. The longest mutation chain that was produced by MCTS was 5; as such, we limit the Random Search to mutate the malware a maximum of 5 times for our comparison. The reasons will become apparent soon through the rest of this section. 

\subsection{Search Results}

\begin{figure}[htb]
\vskip 0.2in
\begin{center}
\centerline{\includegraphics[width=\columnwidth]{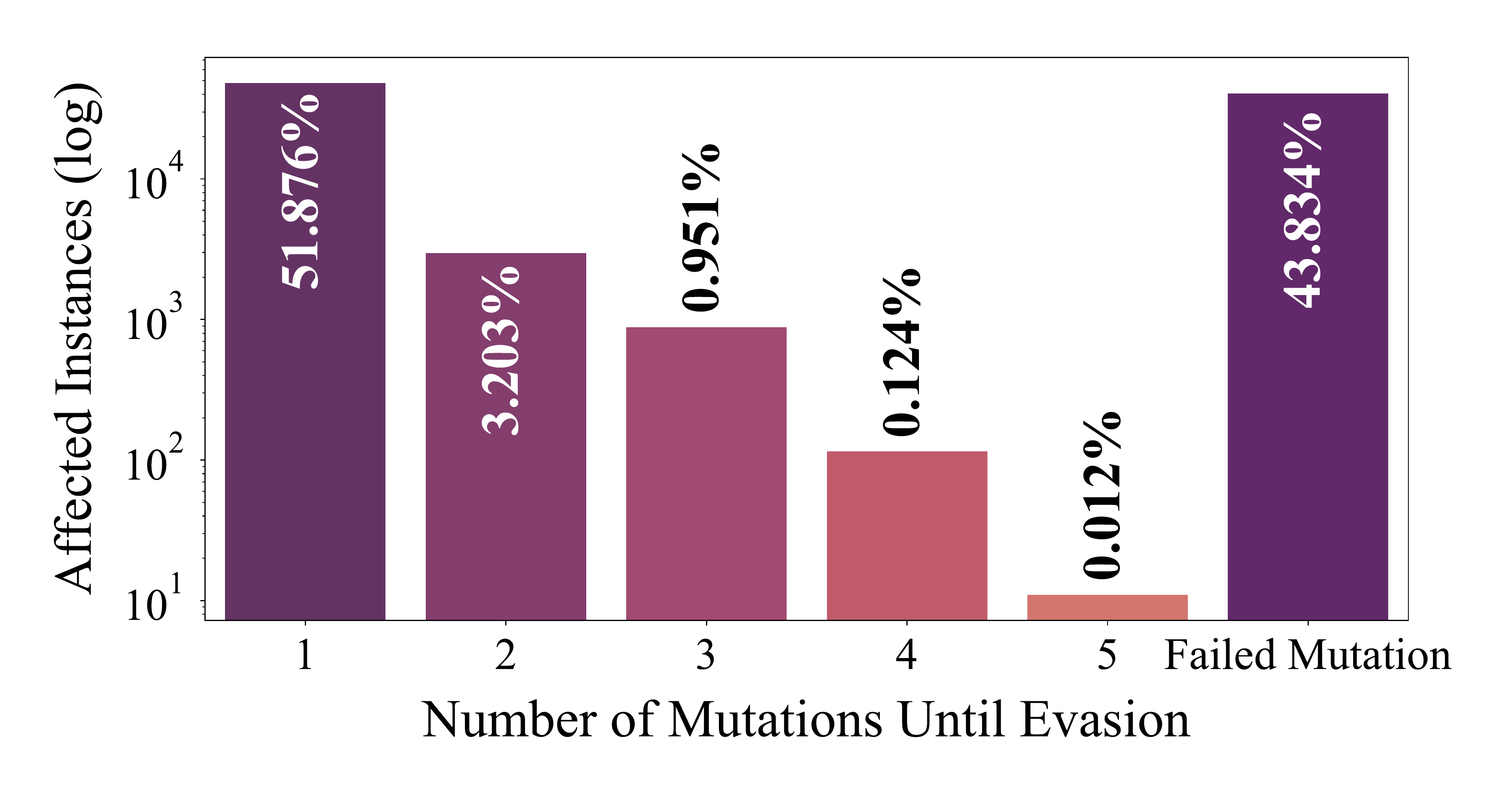}}
\caption{Distribution of number of mutations needed for misclassification over the EMBER test set while using \textbf{MCTS} on the surrogate model. A sample failed mutation, when the algorithm was not able to find a mutation in the given setup/time (for example number of iterations limit is reached).}
\label{fig:mutation_count} %2
\end{center}
\vskip -0.2in
\end{figure}

We begin by introducing the feature changes discovered by the MCTS. Using the surrogate Decision Tree, MCTS was able to find successful mutations for over
56\% of the total malware samples in the EMBER-2018 test set. This is shown in Figure \ref{fig:mutation_count} with the distribution of a total number of
mutations needed for samples to be misclassified. If the algorithm was unable to find a successful mutation in the given setup/time, the sample is classified as a "Failed mutation". Around 52\% of the malware needed only a certificate signature change (\textit{Change Signature}) to be  misclassified as shown in Table \ref{table:mutation_stats}. Specifically, changing the signature alone was enough to alter the prediction for 71\% of the successful mutations. This behavior could be an artifact of the binary nature of the Decision Tree that was used as a surrogate model.

\raggedbottom

The second most potent mutation with MCTS is adding a string that would simultaneously modify the string
entropy of the sample, increase the file size, and the number of strings (\textit{Change String Entropy with Size}). This result is not unexpected as it is an aggregate
mutation affecting 3 of the binary features simultaneously. \textit{Change String Entropy with Size} was a desired path for
19\% of the mutations, and 9,724 malware instances were mutated (i.e. successfully
misclassified) with this modification alone.

\begin{figure}[htb]
\vskip 0.2in
\begin{center}
\centerline{\includegraphics[width=\columnwidth]{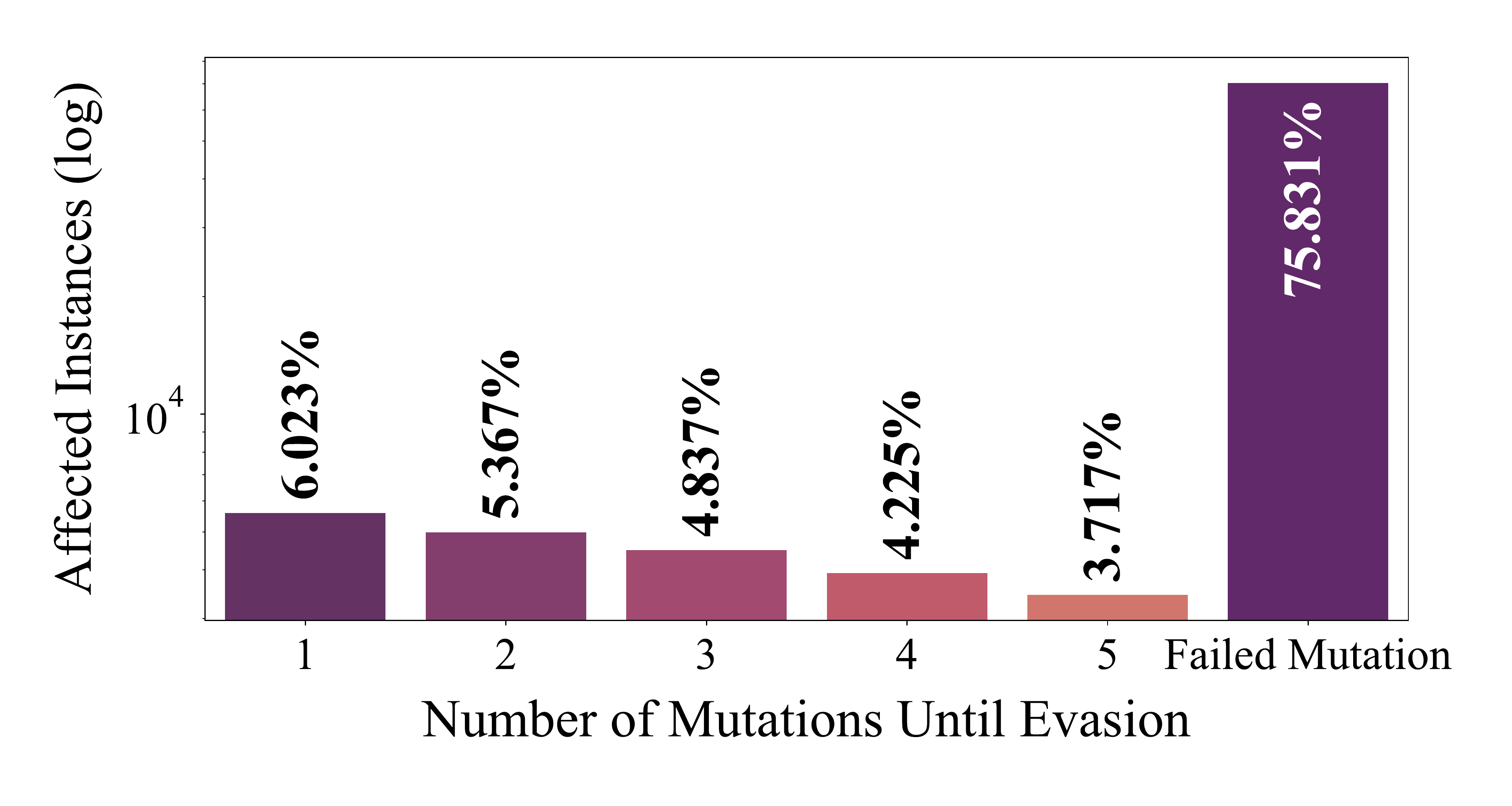}}
\caption{Distribution of the number of mutations needed for misclassification over the EMBER test set while using \textbf{Random Search} on the surrogate model. A sample failed mutation, when the algorithm was not able to find a mutation in the given setup/time (for example number of iterations limit is reached).}
\label{fig:mutation_count_rs} %3
\end{center}
\vskip -0.2in
\end{figure}
\raggedbottom

When using Random Search, the feature that was used the most was \textit{Change Signature}, and the distribution of the number of mutations needed until evasion can be seen in Figure \ref{fig:mutation_count_rs}. Random Search mutated significantly fewer malware samples. In the times where the mutation was successful, it required a higher number of changes.

%\textit{Change Signature} and \textit{Change String Entropy with Size} play a
%big role when using the Random Search as well. However, we see a more balanced
%utilization of all the mutation types, implying that our tree search was able
%to identify a more diverse set of paths. Monte Carlo found 93 unique mutation paths using the MCTS as the verifier, and 61 paths with the Random Search. 

\subsection{Classifier Evasion Results}

\begin{figure}[htb]
\vskip 0.2in
\begin{center}
\centerline{\includegraphics[width=\columnwidth]{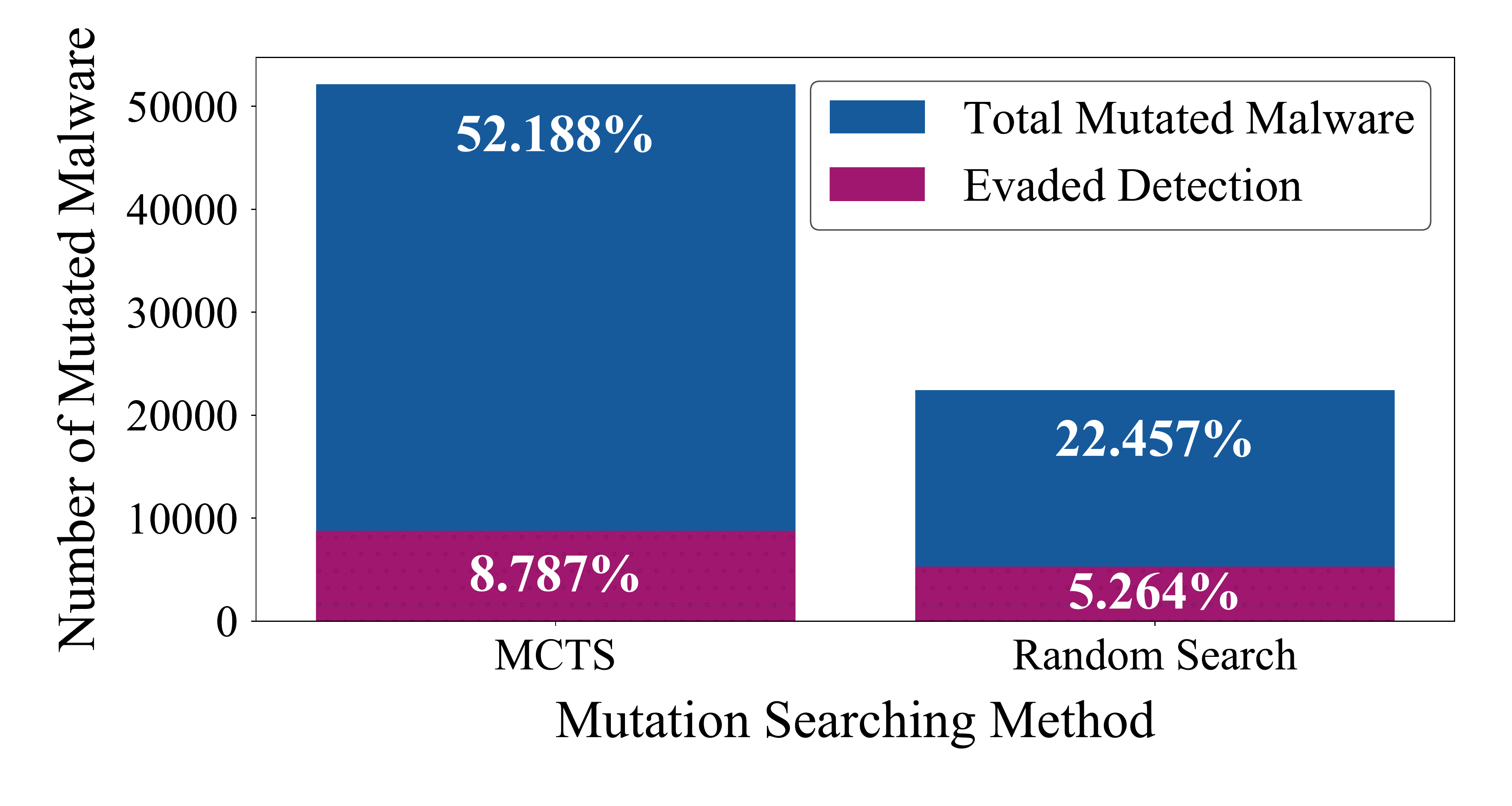}}
\caption{Mutation and evasion results from MCTS and Random Search. \textbf{Blue:} The amount of successfully mutated malware on the surrogate. Failed mutations are excluded. \textbf{Purple (with hatch):} Successful evasions, over the total malware, against the victim.}
\label{fig:evasion_stats_mutated}
\end{center}
\vskip -0.2in
\end{figure}

The performance of the mutations is evaluated against the target API, or victim model, after the modifications found by MCTS and Random Search with the Decision Tree are serialized to mutate the malware samples in the EMBER-2018 test set. The mutated malware is then transformed using the pre-processing pipeline introduced in Section \ref{sec:preprocessing}. Finally, the victim classifier predicts the new labels, and the results from the MCTS and Random Search are compared.

%Figure \ref{fig:roc-auc} shows the performance of both the surrogate and victim model before the mutations. The surrogate Decision Tree F1 score drops from 0.916 to 0.531 for the attacker that is performing the search. Meanwhile, the mutated EMBER-2018 test set drops the victim model's F1 score by 0.048. However, the ROC and PR curves take
%into account all of the benign samples. Therefore, we also show the percentage of the mutated malware that evaded detection in Figure \ref{fig:evasion_stats_mutated} to better understand the performance decay.

%Without the mutations, the victim model correctly predicts 93.32\% of the malicious samples. Utilizing the Decision Tree with Monte Carlo drops this score to 84.78\%. With the Random Search mutations as the validator, we acquire a higher drop of 15.67\% accuracy. This figure takes into account the skipped samples (a malware that was falsely classified prior to mutations are not considered during the evasive feature search) and the samples where successful mutation was not found; therefore, we need to analyze only the samples that were actually modified and how many of those were predicted as benign by the victim API. 

%The mutations produced by MCTS perform better against the victim model. 

As seen in Figure \ref{fig:evasion_stats_mutated}, approximately 8.79\% of the mutations
found by MCTS managed to evade the model. While the majority of the malware is still detected, the increased number of malware passing through the security parameters carry high risk. In comparison, 5.26\% of the mutations found by Random Search evaded detection, and it was able to mutate less than 23\% of the samples. The reduced mutation rate of Random Search is an artifact of limiting the search space to the maximum number of changes performed by MCTS in order to create a more equal comparison. If we were to allow Random Search an unlimited search space, we would be measuring the number of mutations required before miss-classification, instead of finding targeted mutations. In theory, if we were to apply an infinite number of mutations, malware would eventually evade the classifier. However, this would also be detrimental to the functionality of the sample. 

These results show MCTS being able to find more mutations and yield a higher evasion rate in comparison to Random Search in our setup. It should also be noted that our Monte Carlo implementation is flexible for more targeted settings. If we were to narrow down our samples and adjust the various Monte Carlo search hyper-parameters, we could potentially get better results even with the Decision Tree as the surrogate model. In addition, our system is flexible in what model is used as the surrogate. Therefore, an attacker may consider using a more complex surrogate model like a deep learning architecture, or an ensemble of models.

Our intuition, along with the results we have seen in this paper, points to the surrogate model playing a considerable role in the performance of the search. Specifically, we are using a binary decision tree as our surrogate which will need to grow in depth in order to properly train over the dataset. That makes depth-first searches more likely to find and exploit the decision tree, whereas a breadth-favoring approach might work better against an MLP model. Furthermore, the fact that a completely binary mutation was the most used one in both searches might also be an artifact of the binary decision tree. Therefore, the future work can include examining how different surrogate models affect the search as well as experimenting with favoring mutations of slightly longer length in MCTS.

\section*{Conclusion}
Growing popularity of ML based malware detection makes the analysis of these systems against evasive attacks an essential part of cyber defense. We show that a malicious actor can reduce the detection rate of malware samples without the knowledge of the output labels of the Machine Learning model that the adversary camouflages against. Utilizing the adjustable Monte Carlo tree search, with the custom rule set, an adversary can discover the modifications that makes the malware undetectable by checking the changes against a surrogate verifier.

Future work can consider model extraction attacks to create a copy surrogate model to be used as the validator. This can include the \textit{re-training attack} and the \textit{equation-solving attack}. Another direction for future work is to perform direct modifications on the real malicious binaries and verify their functionality after feature alteration(s) via a sandbox setup. Finally, it would also be interesting to try a black box setup where the adversary does not know the features used during the training of the target.

\bibliographystyle{icml2020}
\bibliography{local_references}

\end{document}